\documentclass[conference]{IEEEtran}
\usepackage{graphicx,bm,amsmath,amsfonts}

\newcommand{\lb}{\left}
\newcommand{\rb}{\right}

\newcommand{\Real}{\mathbb{R}}
\newcommand{\Epc}{\mathbb{E}}
\newcommand{\Natural}{\mathbb{N}}

\newcommand{\Tr}{\mathop{\mathrm{Tr}}}
\newcommand{\extr}{\mathop{\mathrm{Extr}}}

\newcommand{\Rr}{\bm{R}_{\mathrm{r}}}
\newcommand{\Rt}{\bm{R}_{\mathrm{t}}}

\begin{document}

\title{A study of the universal threshold in the $\ell_1$ recovery 
 by statistical mechanics}
\author{\IEEEauthorblockN{Koujin Takeda}
\IEEEauthorblockA{Department of Computational Intelligence and\\ 
Systems Science\\ 
Tokyo Institute of Technology\\ 
Yokohama 226-8502, Japan\\
E-mail: takeda@sp.dis.titech.ac.jp
}
\and
\IEEEauthorblockN{Yoshiyuki Kabashima}
\IEEEauthorblockA{Department of Computational Intelligence and\\ 
Systems Science\\ 
Tokyo Institute of Technology\\ 
Yokohama 226-8502, Japan\\
E-mail: kaba@dis.titech.ac.jp
}
}

\maketitle

\begin{abstract}
We discuss the universality of the $\ell_1$ recovery threshold in
compressed sensing. Previous studies in the fields of statistical
mechanics and random matrix integration have shown that 
$\ell_1$ recovery under a random matrix with orthogonal symmetry
has a universal threshold. This indicates that the threshold of
$\ell_1$ recovery under a non-orthogonal random matrix differs
from the universal one. Taking this into account, we use a simple 
random matrix without orthogonal symmetry, where the random entries
are not independent, and show analytically that the threshold of
$\ell_1$ recovery for such a matrix does not coincide with the
universal one. The results of an extensive numerical 
experiment are in good agreement with the analytical results,
which validates our methodology. Though our analysis is based on
replica heuristics in statistical mechanics and is not rigorous,
the findings nevertheless support the fact that the universality 
of the threshold is strongly related to the symmetry of the random matrix.
\end{abstract}

\IEEEpeerreviewmaketitle

\section{Introduction}
\label{sec:sec1}

Compressed sensing is nowadays one of the main topics in information
science, where the sparsity of the signal plays an essential role.
Compressed sensing has been intensively investigated from the
theoretical point of view, and its application has been 
attempted in various fields of engineering.

We start with the basic $\ell_1$-norm recovery problem proposed and
analyzed elsewhere \cite{CRT2006,D2006-1,CT2006}.
The sensing process in this problem is described by a linear equation: 
\begin{equation} 
\bm y = \bm F \bm x^{0}. 
\end{equation} 
(Bold letters denote vectors and matrices.) $\bm x^{0} \in \Real^{N}$ is
the input signal vector, $\bm y\in \Real^{P}$ is a $P$-dimensional
observed signal vector, and $\bm F$ is a $P$-by-$N$ sensing matrix.
We assume that the entries in $\bm F$ are randomly generated.
We define compression rate $\alpha := P/N< 1 $, which is needed for the 
discussion of recovery performance. We also assume that input $\bm x^0$
is also random and that it is drawn from a sparse distribution:
\begin{equation} 
\label{eq:P0dist} 
P(x^0_i) = (1 - \rho) \delta(x^0_i)
+ \rho \widetilde{P}(x^0_i).
\end{equation} 
Parameter $\rho$ is the density of non-zero inputs, which is needed
in the following. The distribution of nonzero entry
$\widetilde{P}(x_i^0)$ can be set arbitrarily in principle. 
Here we set it as Gaussian with zero mean and unit variance.

Within this framework, we consider the $\ell_p$-norm minimization
problem with constraint
\begin{equation} 
\label{eq:l1reconst}
 {\rm minimize} \parallel \!\! \bm x \!\! \parallel_p \ \
 {\rm subject \ to \ \ } \bm y(=\bm F {\bm x}^0)= \bm F \bm x, 
\end{equation} 
where $\parallel \!\! \bm x \!\! \parallel_p
 = \lim_{\epsilon \rightarrow + 0} \sum_{i} |x_i|^{p+\epsilon}$.
In particular we focus on the $\ell_1$ problem ($p=1$). This equation 
offers an algorithm of recovery for original input $\bm x^0$, which
we call $\ell_1$-norm reconstruction. The problem discussed throughout
this paper is a basic question: in which case does the solution
vector of (\ref{eq:l1reconst}) coincide with original input $\bm x_0$?

Under the ansatz of random sensing matrix $\bm F$, the performance of
$\ell_1$-norm recovery has been evaluated using various approaches.
One study using the restricted isometry property \cite{CRT2006}
in conjunction with the large deviation theory of
random matrix spectral edge \cite{CT2006} showed that there is a perfect 
recovery region on the $(\alpha,\rho)$ plane.
Another study using analysis of random polytope projection obtained
a typical reconstruction threshold \cite{DT2005,D2006-2,DT2009-1} that
is in excellent agreement with the boundary between the success and
failure regions obtained in an $\ell_1$-norm reconstruction experiment.
This typical threshold (termed weak threshold elsewhere
\cite{DT2005,D2006-2,DT2009-1}) can 
also be obtained by statistical mechanical analysis based on the
replica method \cite{KWT}, which yields exactly the same analytical
expression for the recovery threshold.

Though the rigorousness of the replica method has not yet been proven
and this method is still a heuristics, it has a significant advantage:
using the replica method, we can analyze problems more general than
the basic $\ell_1$-norm problem. For example, we previously used it
to analyze the correlated sensing problem \cite{TK2010,TK2011}. For 
other generalizations, see the references in \cite{TK2011}.

In this article we focus on the universality of the $\ell_1$-norm
recovery threshold discussed in \cite{DT2009-2}. From the statistical
mechanical point of view, this universality can be comprehended from
rotational symmetry in the matrix integration approach as analyzed
and elucidated elsewhere \cite{KWT}, and with this knowledge we 
can construct a model that breaks such symmetry. In the following
we propose a model of symmetry breaking by introducing a
blockwise-correlated random sensing matrix and give an analytical
expression for the recovery threshold. Using this expression, 
we can qualitatively trace the deformation of the universal recovery
threshold by introducing random matrix correlation.
We also report the results of a numerical experiment for which the
results are in excellent agreement with those of the proposed model.

This article is organized as follows. First we give an overview of
statistical mechanical analysis using a replica method.
Next we address the relationship between this analysis 
and threshold universality from the perspective of matrix integration.
Then, as an example of how statistical mechanics can be used to analyze
problems more general than the basic $\ell_1$-norm problem,
we use it to investigate the deformed problem of an i.i.d. random
matrix.
We next present a blockwise model for observing in detail
the deviation from universality. 
Then we describe the numerical experiment we conducted to verify
the results of the replica method, which lacks rigorousness.
We conclude with a summary of the key points and a short discussion.

\section{Overview of statistical mechanical analysis}
\label{sec:sec2}

We start with an outline of statistical mechanical analysis,
as proposed elsewhere \cite{KWT}. We focus on the basic model,
where each entry in matrix $\bm F$ is drawn from a Gaussian 
distribution with zero mean and variance $N^{-1}$. 
For the moment we do not restrict ourselves to the $\ell_1$-norm
problem and consider instead the $\ell_p$-norm problem.

The first step of the analysis is to define quantity $C_p$:
\begin{eqnarray}
 C_p &:=&
 -\lim_{\beta \rightarrow \infty} \lim_{N \rightarrow \infty}
 \frac{1}{\beta N} \nonumber \\
 && \times \ln \int d \bm x \exp (- \beta \parallel 
 \!\! \bm x \!\! \parallel _p) 
 \delta (\bm F (\bm x^a - \bm x^0)).
\end{eqnarray}
This definition describes the minimized $\ell_p$-norm
(divided by $N$) under the condition
$\bm y (= \bm F \bm x_0)= \bm F \bm x$, which is clearly obtained 
by taking the limit of $\beta \rightarrow \infty$.
In the present case, matrix $\bm F$ and input $\bm x_0$ are random,
and we need to take the average w.r.t. them. This requires 
calculating the average of the logarithmic quantity on the rhs,
which is an obstacle to the analysis. To overcome this obstacle,
we resort to the replica method, which has not been shown to be
rigorous but heuristically gives the exact result. With the replica 
method, $C_p$ after averaging is 
\begin{eqnarray} 
\label{eq:freeenergy}
 \Epc [ C_p ]_{\bm F, \bm x_0}
 & \!\!\!\!\! := \!\!\!\!\! & \!\!\! 
 - \!\! \lim_{\beta \rightarrow \infty}
 \lim_{N \rightarrow \infty} \frac{1}{\beta N}
 \lim_{n \rightarrow 0} \frac{\partial}{\partial n}
 \ln \Epc [ Z^n (\bm F, \bm x^0)]_{\bm F, \bm x^0}, 
 \nonumber \\
\end{eqnarray} 
where $\Epc [\ ]_{\bm X}$ denotes the average w.r.t. variable
(vector, matrix) $\bm X$; the $n$th power of factor $Z(\bm F, \bm x^0)$ is 
\begin{eqnarray} 
\label{eq:nthmoment}
 Z^{n}(\bm F, \bm x^0)
 & \hspace{-0.3cm} := & \hspace{-0.2cm}
 \prod_{a=1}^{n} \int {d \bm x^a} \exp (- \beta \parallel 
 \!\! \bm x^a \!\! \parallel _p)
 \delta (\bm F (\bm x^a - \bm x^0)) \nonumber \\
 && \hspace{-2.0cm} = \prod_{a=1}^{n} \int {d \bm x^a}
 \lim_{\tau \rightarrow +0} \frac{1}{(\sqrt{2 \pi \tau})^{nP}} \nonumber \\
 && \hspace{-2.5cm} \times \exp
 \left[ - \sum_{a=1}^{p} \beta \parallel \!\! \bm x^a \!\! 
 \parallel _p -\frac{1}{2 \tau} \sum_{a=1}^{n}
 (\bm x^a - \bm x^0)^{T} \bm F^{T} \bm F
 (\bm  x^a - \bm x^0) \right]. \nonumber \\
\end{eqnarray}
This means that we can estimate the logarithmic quantity from
the positive integer moment with identity
$\Epc [\ln X] = \lim_{n \rightarrow 0} 
 \partial \ln \Epc [X^n] / \partial n$. 
Superscript $a$ on $\bm x$ denotes the ``replica'' number
introduced for estimating $\Epc [Z^n (\bm F, \bm x^0)]_{\bm F, \bm x^0}$.
After the average is taken over Gaussian random matrix
$\bm F$ and limit $\tau \rightarrow +0$,
\begin{eqnarray} 
\label{eq:nthave}
 \hspace{0cm} \Epc[Z^n (\bm F, \bm x^0)]_{\bm F} 
 \!\!\!\! &=& \!\!\!\!  \int d \bm x^0
 \int d\bm Q  \prod_{a=1}^{n} \int d \bm x^{a} \nonumber \\
 && \hspace{-3.0cm}
 \times
 \exp \lb( -\frac{ \alpha N}{2}
 \Tr \ln \bm S  - \sum_{a=1}^{n} \beta \parallel \!\!
 \bm x^a \!\! \parallel _p \rb)  \Pi^{(n)}(\bm Q, \bm x^a).
\end{eqnarray} 
We omit the trivial overall factor as it is irrelevant to the analysis.
Square matrix $(\bm S)_{ab} := Q_{ab} - 2 Q_{0a} + \rho$ is
$n$-dimensional; $Q_{ab}$ is defined in $\Pi^{(n)} (\bm Q, \bm x^a)$.
Delta function constraint $\Pi^{(n)} (\bm Q, \bm x^a)$ is given as
\begin{eqnarray} 
\label{eq:defPi}
 \Pi^{(n)}(\bm Q, \bm x^a) 
 &:=& 
 \nonumber \\
 && \hspace{-2.2cm} \prod_{a=1}^{n} \int^{+i\infty}_{-i\infty}
 d \widetilde{Q}_{aa} 
 \exp \left\{ N \widetilde{Q}_{aa} (\bm x^{a T} \bm x^{a}-NQ_{aa}) \right\} 
 \nonumber \\
 && \hspace{-2.5cm} \times \prod_{a<b} \int^{+i\infty}_{-i\infty}
 d \widetilde{Q}_{ab}
 \exp \left\{ N \widetilde{Q}_{ab} (\bm x^{a T} \bm x^{b} -NQ_{ab}) \right\} 
 \nonumber \\
 && \hspace{-2.5cm} \times \prod_{a=1}^{n} \int^{+i\infty}_{-i\infty}
 d \widetilde{Q}_{0a}
 \exp \left\{ N \widetilde{Q}_{0a} 
 ( \bm x^{a T} \bm x^{0}-NQ_{0a} ) \right\}.
\end{eqnarray} 
In this definition, dual matrix $\widetilde{\bm Q}$ is introduced
as a collection of integration variables for Fourier representation
of the delta function.

Computing the $n\rightarrow 0$ limit requires analytic continuation
from $n \in \Natural$ to $n \in \Real$. To achieve this,
we follow the standard procedure in the replica method and assume
replica symmetry regarding matrices $\bm Q$ and $\widetilde{\bm 
Q}$. Let $q=Q_{ab}, \widetilde{q}=\widetilde{Q}_{ab}$
(both for $a \ne b$), $Q = Q_{aa}, \widetilde{Q} = \widetilde{Q}_{aa}$,
$m=Q_{a0}$ and $\widetilde{m} = \tilde{Q}_{a0}$, yielding
$S_{aa} = Q-2m+u$ and $S_{ab}=q-2m+u \ (a \ne b)$.
By diagonalization of matrix $\bm S$, $\Tr \ln \bm S$ is evaluated as
\begin{eqnarray} 
\label{eq:Trln}
 \Tr \ln \bm S \!\!\!\!\! &=& \!\!\!\!
 (n-1) \ln (Q-q) + \ln \left\{ Q -q + n( q -2m + \rho) \right\}.
\nonumber \\
\end{eqnarray} 
Under the assumption of replica symmetry, the $\bm x^a$-dependent
part, namely the $\ell_p$-norm and $\Pi^{(n)}(\bm Q, \bm x^a)$ are
deformed to
\begin{eqnarray} 
\label{eq:Sn}
 && \hspace{-0.7cm} 
 \prod_{a=1}^{n} \int d {\bm x}^{a}
 \exp \lb( - \sum_{a=1}^{n} \beta \parallel \!\!
 \bm x^a \!\! \parallel _p \rb)
 \Pi^{(n)}(\bm Q, \bm x^a) \nonumber \\
 &=&
 \exp \lb( -NnQ\widetilde{Q} - N\frac{n(n-1)}{2}q \widetilde{q}
 -Nnm\widetilde{m} \rb) \nonumber \\
 && \times \left. \int D \widetilde{z}
 \left( \int d x \exp \lb( N \lb\{ \left( \widetilde{Q} - 
 \frac{\widetilde{q}}{2} \right) x^2
 \right. \right. \right. \right. \nonumber \\
 && \left. \left. \left. \hspace{1cm}
 + x^{T} \left( \widetilde{m} x^{0}
 + \sqrt{\widetilde{q}} \widetilde{z} \right)
 - \beta | x |^p \right\} \rb] 
 \right)^{n}\!\!\!, 
\end{eqnarray} 
where $D \widetilde{z} := (\sqrt{2\pi})^{-1} \int^{\infty}_{-\infty}
d \widetilde{z} e^{- \widetilde{z}^2 /2}$. In the last line,
the interaction between replicas (namely $\bm x^{a T} \bm x^b$)
is removed by incorporating auxiliary Gaussian variable 
$\widetilde{z}$ (often called Hubbard-Stratonovich transformation
in physics) and decomposing all replicas.

From (\ref{eq:Trln}, \ref{eq:Sn}) we find that all $n$-dependent
factors are taken as defined for $n \in \Real$ (putting mathematical
rigorousness aside), which allows us to calculate $n \rightarrow 0$ limit.
For convenience of further analysis, we redefine the auxiliary
variables, $\widehat{m} := \beta^{-1} \widetilde{m}, 
\widehat{\chi} := \beta^{-2} \widetilde{q}, \chi := \beta (Q - q),$
and $\widehat{Q} := \beta^{-1} (-2\widetilde{Q} + \widetilde{q})$,
and also introduce the function 
\begin{equation} 
 \phi_p(h,\widehat{Q}) \hspace{-0.1cm} := \hspace{-0.1cm}
 \frac{1}{N} {\rm min}_{x}
 \left\{ \frac{\widehat{Q}}{2}
 x^2 -  h x + |x|^p \right\}. 
\end{equation} 
These variables and function are used to simplify the factor in
$\int D \widetilde{z}$ in (\ref{eq:Sn}) to
$\exp \{ \! -\beta N n \phi_{p}( \widehat{m} x^{0} + 
\sqrt{\widehat{\chi}} \widetilde{z}, \widehat{Q}) \! \}$
for $\beta \rightarrow \infty$. 

After combining these results and computing the average w.r.t. $\bm x_0$,
we compute the limit $N \rightarrow \infty$. As a result,
a six-dimensional integral w.r.t. ${\widehat{Q},\widehat{m},
\widehat{\chi},q,m,\chi}$ is replaced with one w.r.t. their 
extremal values by asymptotic analysis.
This integral is denoted by the symbol $\extr$ in the following.
(Although the commutativity of the limits $n \rightarrow 0$ and $N 
\rightarrow \infty$ has not been shown, this has not been a concern
in standard replica analysis.) Finally, after computing
the limit $n \rightarrow 0$, we arrive at the final expression:
\begin{eqnarray} 
\label{eq:Cpfinal}
 \Epc[C_p]_{\bm F,\bm x^0} 
 &\hspace{-0.3cm} =& \hspace{-0.3cm} 
 \extr_{\substack{\widehat{Q},\widehat{m},\widehat{\chi} \\ q,m,\chi}}
 \! \left\{ \! \left. \frac{ \alpha ( q -2m + u)}
 {2 \chi} \! + \! \left( \! \frac{\chi \widehat{\chi}}{2}
 - \frac{q \widehat{Q}}{2} + m \widehat{m} 
 \! \right) \right. \right. \nonumber \\
 && \hspace{-0.3cm} + \!\! \left. 
 \int d x^0 P(x^0) \int D \tilde{\bm z} 
 \phi_{p}( \widehat{m} {x}^{0}
 + \sqrt{\widehat{\chi}} \tilde{z}, 
 \widehat{Q})  \! \right\}\!\! . 
\end{eqnarray} 
For evaluation of the threshold, we can extract some information from
(\ref{eq:Cpfinal}). 
As shown in (\ref{eq:Cpfinal}), $\Epc [C_p]_{\bm F, \bm x^0}$ is nothing but 
the minimized $\ell_p$-norm after averaging,
which can be calculated using the extremal values of the six variables.
Returning to the definition of $q,m$ (delta function in 
(\ref{eq:defPi})) and remembering that $\bm x$ is the result of
recovery and $\bm x^0$ is the original input,
we can see that $q=m=\rho$ must be satisfied at the extremal of 
$q,m$ when the recovery is successful, whereas $q \ne m$ is expected
in the unsuccessful case. For threshold evaluation, we need to observe
the bifurcation from $q=m$ to $q \ne m$. This statement assumes
continuous bifurcation, or second-order phase transition in the context
of statistical mechanics, which is true for the present problem.

For observing bifurcation, it is more convenient to use
the variable $\chi(=\beta(Q-q))$. In this problem, $Q=q=m$ holds
for successful recovery, while $Q \ne q \ne m$ for failure, 
which means $\chi=0$ and $\chi \ne 0$ for success and failure,
respectively. Henceforth, we use the bifurcation from $\chi=0$ to
$\chi \ne 0$ at the extremal in (\ref{eq:Cpfinal}), which yields
the recovery threshold in conjunction with the conditions 
of the other five variables. For $p=1$ ($\ell_1$-norm),
the threshold can be expressed as simply two equations:
\begin{eqnarray}
\label{eq:L1threshold}
 \alpha &=& 2 (1 - \rho) H \left( 1 / \sqrt{\widehat{\chi}} \right) + \rho,
 \nonumber \\
 \widehat{\chi} &=&
 \alpha^{-1} \left\{ 2 (1 - \rho) \left(
 (\widehat{\chi} +1 ) H \left( 1 / \sqrt{\widehat{\chi}} \right)
 \right. \right.
 \nonumber \\
 && \left. \left. \hspace{0.5cm} 
 - (2 \pi)^{-1/2} 
 \sqrt{\widehat{\chi}} e^{-1/2\widehat{\chi}}
  +\rho(\widehat{\chi}+1)
 \right)
 \right\},
\end{eqnarray}
where $H(x) := (2 \pi)^{-1/2} \int_x^{\infty} dt e^{-t^2/2}$ is
a complementary error function (slightly different definition
from the standard), and elimination of $\widehat{\chi}$ yields
the relation between $\alpha$ and $\rho$, which is the 
$\ell_1$ recovery threshold. This result coincides with the weak
threshold in \cite{DT2005,D2006-2,DT2009-1}, computed from random
polytope projection. (The equivalence is noted in \cite{KWT}.
Actually, the extremal condition w.r.t. $\nu$ for cross-polytope
in Section 6.2 in \cite{DT2009-1} is shown to be the same
as (\ref{eq:L1threshold}) after some algebra.)

\section{Universality from statistical mechanical analysis: 
matrix integration}
\label{sec:sec3}

Numerical investigation of the universality of the $\ell_1$-norm
recovery threshold \cite{DT2009-2} using several kinds of i.i.d.
random entries in $\bm F$, such as Gaussian and Bernoulli, and
several random orthogonal bases, such as Fourier and Hadamard, 
indicated that the threshold under random matrices is universal. 

From a statistical mechanical point of view, this universality is
understood by the matrix integration formula \cite{KWT}.
Here we use the formula from Lie group theory \cite{HC},
which is equivalent to one from mathematical physics \cite{IZ}, 
called the Harish-Chandra-Itzykson-Zuber integral,
\begin{eqnarray}
\label{eq:HIZ}
 \frac{\int d \bm O \exp \left\{ 
 \frac{1}{2} \Tr \bm O \bm D \bm O^{T} \bm L  \right\} }
 {\int d \bm O}
 = \exp \left\{ N \Tr G \left( \frac{\bm L}{N} \right)
 \right\},
\end{eqnarray}
for computing $N \rightarrow \infty$, where $\bm O$ is an orthogonal,
$\bm D$ is a diagonal, and $\bm L$ is an arbitrary matrix.
(This formula was originally given for unitary matrix integration.)
All matrices are square and $N$-dimensional; $d \bm O$ is the 
Haar measure of the $N$-dimensional orthogonal group.
The function $G$ on the rhs is computed from (see e.g. \cite{MPR,CDL,TUK})
\begin{eqnarray}
 G(x) &=& \frac{1}{2} \int_0^x dt \left( \Lambda(t) - \frac{1}{t} \right),
\end{eqnarray}
which is known as $R$-transformation in free probability theory
\cite{TVbook}. The function $\Lambda(t)$ is implicitly given by
Cauchy (or Stieltjes) transformation,
\begin{eqnarray}
 x &=& \int d \lambda \frac{\rho_{\bm D}(\lambda)}{\Lambda (x) - \lambda},
\end{eqnarray}
where $\rho_{\bm D}(\lambda)$ is the density of the diagonal element
values in $\bm D$.

The random matrix ensemble (Wishart ensemble) $\bm F^{T} \bm F$,
where $\bm F$ is a $P$-by-$N$ i.i.d. Gaussian random matrix with
variance $N^{-1}$, is assumed to be equivalent to the ensemble
$\bm O \bm D \bm O^{T}$ generated by arbitrary orthogonal 
matrix $\bm O$ under the condition that $\rho_{\bm D}(\lambda)$ follows 
Mar\u{c}enko-Pastur law \cite{MPlaw} for $\alpha=P/N<1$:
\begin{eqnarray}
\label{eq:MPdis}
 \rho_{\bm D}(\lambda) & = &
 \left( 1 - \alpha \right) \delta(\lambda) \nonumber \\
 && \hspace{-1.8cm}
 + \frac{1}{2 \pi}
 \frac{\sqrt{ (\lambda_+ - \lambda )( \lambda - \lambda_-)} }
 {\lambda}
 \Theta(\lambda_+ - \lambda) \Theta(\lambda - \lambda_-),
\end{eqnarray}
where $\lambda_{\pm} = ( 1 \pm \alpha^{1/2} )^2$ and $\Theta(x)$
is a Heaviside function. (For a unitary ensemble, the equivalence
is as shown in \cite{TVbook}.) This is 
known as the asymptotic eigenvalue density of the Wishart random
matrix ensemble $\bm F^{T} \bm F$.

We can apply this formula to the average w.r.t. random matrix $\bm F$ in 
(\ref{eq:nthmoment}). For (\ref{eq:MPdis}), the function $G(x)$ is
computed as $G(x) = - (\alpha / 2) \ln (1 - x )$,
and applying (\ref{eq:HIZ}) to (\ref{eq:nthmoment})
(integration performed over Haar measure $d \bm O$) results in an 
average the same as that obtained using (\ref{eq:nthave}),
as noted in \cite{KWT}.

This strongly suggests that rotational invariance of the random
matrix ensemble $\bm F^T \bm F$ is combined with the universal
threshold because the result of analysis using matrix 
integration w.r.t. the orthogonal group Haar measure is the same as
that using integration of a Gaussian random matrix with i.i.d. entries
performed using (\ref{eq:nthmoment}). This implies that universality
breakdown requires a random matrix ensemble that breaks such 
symmetry. In the following, we present a symmetry breaking model
and see how the threshold deviates from the universal one.

\section{Deformed problem}
\label{sec:sec4}

As mentioned above, statistical mechanics can be used to analyze
problems more general than the basic $\ell_1$-norm problem.
For example, we previously used it to investigate 
the deformed problem of an i.i.d. random matrix \cite{TK2010},
where $\bm F$ is given by
\begin{equation} 
\label{eq:defF}
 \bm F = \sqrt{\Rr} \bm \Xi \sqrt{\Rt}.
\end{equation} 
Matrices $\Rr$ and $\Rt$ are respectively $P$- and $N$-dimensional
deterministic square symmetric matrices. The square root of square matrix
$\bm A$ is defined by
$\bm A = \sqrt{\bm A}^{T} \sqrt{\bm A}$ 
(e.g. the Cholesky decomposition can be used for positive-definite
 $\bm A$.).
The $\bm \Xi$ is a 
$P$-by-$N$ rectangular matrix with entries that are i.i.d.
Gaussian random variables with zero mean and variance $N^{-1}$.
As stated elsewhere \cite{TK2010}, $\Rr$ and $\Rt$ respectively
describe the correlation among observation vectors and the correlation 
among the representation bases of the sparse input signals.
(Such a framework is described elsewhere \cite{CW2008}.)
When $\Rr$ and $\Rt$ are identities, $\bm F$ becomes an 
i.i.d. random matrix, and the problem returns to the original one.

We previously applied the replica method to this problem and computed
$\Epc [C_p]_{\bm F, \bm x^0}$ \cite{TK2010}. The result was
\begin{eqnarray} 
\label{eq:Cpdeformfinal}
 \Epc [C_p]_{\bm F, \bm x^0} 
 \!\!\!\!\! &=& \!\!\!\!\!
 \extr_{\substack{\widehat{Q},\widehat{m},\widehat{\chi} \\ q,m,\chi}}
 \left( \left. \! \frac{ \alpha ( q -2m + u)}{2 \chi}
 + \! \left( \frac{\chi \widehat{\chi}}{2} - \frac{q \widehat{Q}}{2}
 + m \widehat{m} \! \right) \right. \right. \nonumber \\
 && \hspace{-2cm} + \left. \left\{ \prod_{i} \int d x_i^0 P(x_i^0) 
 \int D \tilde{\bm z} \widetilde{\phi}_{p}( \widehat{m} \sqrt{\Rt}
 {\bm x}^{0} + \sqrt{\widehat{\chi}} \tilde{\bm z}, 
 \widehat{Q}) \right\} \right), \nonumber \\
\end{eqnarray} 
where $\widetilde{\phi}_p(\bm h, \widehat{Q})$ is given as the solution
to an $N$-variable minimization problem as
\begin{equation} 
\label{eq:Phideform}
 \widetilde{\phi}_p(\bm h,\widehat{Q})
 \hspace{-0.1cm} := \hspace{-0.1cm} \frac{1}{N} 
 {\rm min}_{\bm x} \left\{ \frac{\widehat{Q}}{2} \bm x^{T} \Rt 
 \bm x - \bm h^{T} \sqrt{\Rt} \bm x + \parallel \!\! \bm x\!\! \parallel
 _{p} \right\}. 
\end{equation}
The result clearly does not depend on $\Rr$
(assuming $\Rr$ is full-rank), and does
only on $\Rt$. This can be understood from matrix integration.
Suppose that $\Rt$ is an identity matrix; the ensemble
$\bm F^{T} \bm F = \bm \Xi^T \bm \Rr \bm \Xi $
is then equivalent to $\bm O \bm D \bm O^{T}$ in the previous
section because matrix $\Rr$ can be eliminated by
the redefinition of $\bm \Xi$
(after normalization, which changes the Mar\u{c}enko-Pastur law
of Wishart ensemble $\bm \Xi^T \bm \Xi$ and $G(x)$ in (\ref{eq:HIZ}),
however is irrelevant to the universality \cite{TK2010}).
On the other hand, $\Rt$ cannot be eliminated in the same manner and
affects the Haar measure $d\bm O$.
This implies that such a random matrix ensemble will differ 
from the one connected from 
a diagonal matrix like Mar\u{c}enko-Pastur by orthogonal transformation.

We can calculate the threshold for this deformed problem in a manner
similar to that used in the previous section by investigating
the bifurcation from $\chi=0$ to $\chi \ne 0$. 
Note that, in this problem, we must solve an $N$-variable minimization
problem, as in (\ref{eq:Phideform}),
while in the original problem this is simply a minimization with only 
one variable. Such minimization generally requires a numerical method
(e.g., Monte Carlo) as used previously \cite{TK2010}.

We also studied another deformed problem, where input signal
$\bm x^0$ is sparse and directly correlated \cite{TK2011};
since this is beyond the scope of this article, we omit 
the details here.

\section{Example of non-universal threshold: blockwise model}
\label{sec:sec5}

For observing in detail the deviation from universality
when using the deformed model and statistical mechanics both analytically
and quantitatively, we propose using a blockwise model. 
The sensing matrix in this model is the one given in (\ref{eq:defF}),
and $\Rt$ is composed of $2$-by-$2$ diagonal blocks.
(Input size $N$ is assumed to be even).
\begin{eqnarray}
\label{eq:defRt}
 \Rt &=& \left(
\begin{array}{cc}
 1 & r \\
 r & 1 \\
\end{array}
 \right) \otimes
 \bm I_{N/2}, \nonumber \\
 {\rm accordingly\ }
  \sqrt{\Rt} &=&
 \left(
\begin{array}{cc}
 l_+ & l_- \\
 l_- & l_+ \\
\end{array}
 \right) \otimes
 \bm I_{N/2},
\end{eqnarray} 
where $l_\pm := (\sqrt{1+r} \pm \sqrt{1-r})/2$, and $\bm I_{N/2}$
denotes an identity matrix of size $N/2$.
$\sqrt{\Rt}$ is chosen to be symmetric, and $\Rr = \bm I_P$.
This blockwise matrix represents the case
in which all input signals (or equivalently corresponding
representation bases) are correlated with their partner.

By substituting these signals or bases into (\ref{eq:Cpdeformfinal})
and (\ref{eq:Phideform}) and observing the bifurcation from
$\chi=0$ to $\chi \ne 0$, we can obtain the recovery threshold
for the blockwise model. This analysis scheme is almost the 
same as that previously proposed \cite{TK2010}, so the details are omitted.

The final equations for the threshold are
\begin{eqnarray}
\label{eq:L1deformthreshold}
 \alpha &=& \frac{1}{2 \sqrt{\widehat{\chi}}}
 \int Dz_1 \int Dz_2 V\left(z_1,z_2,\widehat{\chi}\right),
 \nonumber \\
 \widehat{\chi} &=& \frac{1}{2 \alpha}
 \int Dz_1 \int Dz_2 W\left(z_1,z_2,\widehat{\chi}\right),
\end{eqnarray}
which corresponds to (\ref{eq:L1threshold}) for the original basic 
$\ell_1$-norm problem.
The functions $V\left(z_1,z_2,\widehat{\chi}\right)$ and 
$W\left(z_1,z_2,\widehat{\chi}\right)$ are defined as
\begin{eqnarray}
\label{eq:defV}
 && V\left(z_1,z_2,\widehat{\chi}\right) \nonumber \\
 &:=& \hspace{-0.5cm}
 \sum_{\xi_1,\xi_2 = 0,1}
 \left\{ (1-\rho) \delta_{\xi_1,0} + \rho \delta_{\xi_1,1} \right\}
 \left\{ (1-\rho) \delta_{\xi_2,0} + \rho \delta_{\xi_2,1} \right\}
 \nonumber \\
 && \hspace{-0.9cm} \times
 \sum_{ \sigma_1,\sigma_2 = \pm 1} 
 \sum_{i,j=1,2} \frac{1}{4(1-r^2)} \left( \sqrt{\Rt^B} \right)_{ij}
 \nonumber \\
 &&  \hspace{2cm} \times z_i x_{\xi_1,\xi_2}^{(j)} \left(
 \sigma_1,\sigma_2,z_1,z_2,\sqrt{\widehat{\chi}}
 \right), 
\end{eqnarray}
\begin{eqnarray}
\label{eq:defW}
 && W\left(z_1,z_2,\widehat{\chi}\right) \nonumber \\
 &:=& \hspace{-0.5cm}
 \sum_{\xi_1,\xi_2 = 0,1}
 \left\{ (1-\rho) \delta_{\xi_1,0} + \rho \delta_{\xi_1,1} \right\}
 \left\{ (1-\rho) \delta_{\xi_2,0} + \rho \delta_{\xi_2,1} \right\}
 \nonumber \\
 && \hspace{-0.9cm}
 \times
 \sum_{ \sigma_1,\sigma_2 = \pm 1} 
 \sum_{i,j=1,2}  \frac{1}{4(1-r^2)} \left( \Rt^B \right)_{ij}
 \nonumber \\
 && \hspace{-0.9cm}
 \times x_{\xi_1,\xi_2}^{(i)} \!\! \left(
 \sigma_1,\sigma_2,z_1,z_2,\sqrt{\widehat{\chi}}
 \right) \!
 x_{\xi_1,\xi_2}^{(j)} \!\! \left(
 \sigma_1,\sigma_2,z_1,z_2,\sqrt{\widehat{\chi}}
 \right),  
\end{eqnarray}
where $\Rt^B$ and $\sqrt{\Rt^B}$ are the 2-by-2 block
matrices in (\ref{eq:defRt}). Boolean variables $\xi_1$ and $\xi_2$
represent the case in which each input signal is 
respectively zero and nonzero for pairwise input.
The functions $x_{\xi_1,\xi_2}^{(1),(2)}$ 
for each $\xi_1$ and $\xi_2$ are given as
\begin{eqnarray}
 x_{0,0}^{(1)} (z_1,z_2,\widehat{\chi}) \!\!\!\! 
 &:=& \hspace{-0.5cm}
 \sum_{\eta_1,\eta_2=\pm 1} \!\!\!\!\!
 \Omega_{\eta_1} \!\!
 \left( (\widehat{l}_+ z_1 + \widehat{l}_- z_2)
 \sqrt{\widehat{\chi}} + r \eta_2 - \eta_1 \right)
 \nonumber \\
 && \hspace{-0.5cm}
 \times \Theta\left( \eta_2 \left\{
 (\widehat{l}_- z_1 + \widehat{l}_+ z_2 ) \sqrt{\widehat{\chi}} 
 + r \eta_1 - \eta_2 \right\}
 \right)
 \nonumber \\
 && \hspace{-1cm}
 + \sum_{\eta = \pm 1} (1-r^2)\Omega_{\eta}
 \left( (l_+ z_1 + l_- z_2) \sqrt{\widehat{\chi}} - \eta \right)
 \nonumber \\
 && \hspace{-0.5cm}
 \times \Theta\left( 
 (\widehat{l}_- z_1 + \widehat{l}_+ z_2 ) \sqrt{\widehat{\chi}} 
 + r \eta + 1
 \right)
 \nonumber \\
 && \hspace{-0.5cm}
 \times \Theta\left( - \left\{
 (\widehat{l}_- z_1 + \widehat{l}_+ z_2 ) \sqrt{\widehat{\chi}} 
 + r \eta - 1 \right\}
 \right),
 \nonumber \\
 x_{0,0}^{(2)} (z_1,z_2,\widehat{\chi})
 &:=& {\rm replace} \{\widehat{l}_+,\widehat{l}_-,l_+,l_- \} 
 \ {\rm in }  \nonumber \\
 &&   \hspace{-0.5cm} x_{0,0}^{(1)} (z_1,z_2,\widehat{\chi})
 \ {\rm with} \  \{\widehat{l}_-,\widehat{l}_+,l_-,l_+ \},
\end{eqnarray}
\begin{eqnarray}
 x_{0,1}^{(1)} (\sigma_2,z_1,z_2,\widehat{\chi}) 
 \!\!\!\!\! &:=& \!\!\!\!\!\!\!\! \sum_{\eta=\pm 1}
 \! \Omega_{\eta} \!
 \left( (\widehat{l}_+ z_1 + \widehat{l}_- z_2)\sqrt{\widehat{\chi}}
 + r \sigma_2 - \eta \! \right)\!, \nonumber \\
 x_{0,1}^{(2)} (\sigma_2,z_1,z_2,\widehat{\chi}) 
 \!\! &:=& \!\!
 - r x_{0,1}^{(1)} (\sigma,z_1,z_2,\widehat{\chi} ) \nonumber \\
 && \hspace{-1cm}
 + (1-r^2) 
 \left\{ (l_- z_1 + l_+ z_2) \sqrt{\widehat{\chi}} - \sigma_2 \right\},
\end{eqnarray}
\begin{eqnarray}
 x_{1,0}^{(2)} (\sigma_1,z_1,z_2,\widehat{\chi}) 
 \!\!\!\!\! &:=& \!\!\!\!\!\!\!\! \sum_{\eta=\pm 1}  
 \! \Omega_{\eta} \!
 \left( (\widehat{l}_- z_1 + \widehat{l}_+ z_2)\sqrt{\widehat{\chi}}
 + r \sigma_1 - \eta \! \right)\! ,  \nonumber \\
 x_{1,0}^{(1)} (\sigma_1,z_1,z_2,\widehat{\chi}) 
 \!\! &:=& \!\! 
 - r x_{1,0}^{(2)} (\sigma,z_1,z_2,\widehat{\chi} ) \nonumber \\
 && \hspace{-1cm}
 + (1-r^2) \left\{
 (l_+ z_1 + l_- z_2) \sqrt{\widehat{\chi}} - \sigma_1 \right\},
\end{eqnarray}
\begin{eqnarray}
 x_{1,1}^{(1)} (\sigma_1,\sigma_2, z_1,z_2,\widehat{\chi})
 &:=& (\widehat{l}_+ z_1 + \widehat{l}_- z_2)
 \sqrt{\widehat{\chi}} - \sigma_1 + r \sigma_2,  \nonumber \\
 x_{1,1}^{(2)} (\sigma_1,\sigma_2, z_1,z_2,\widehat{\chi})
 &:=& (\widehat{l}_- z_1 + \widehat{l}_+ z_2)
 \sqrt{\widehat{\chi}} - \sigma_2 + r \sigma_1,
 \nonumber \\
\end{eqnarray}
where $\Omega_{\eta} (x) := x \Theta (\eta x)$ for $\eta = \pm 1$ and 
$\widehat{l}_\pm := l_{\pm} - r l_{\mp}$. In some cases, $x_{\xi_1, 
\xi_2}^{(1),(2)}$ does not depend on $\sigma_1$ and/or $\sigma_2$.
Nevertheless, the summations with respect to $\sigma_1$ and $\sigma_2$
in (\ref{eq:defV}) and (\ref{eq:defW}) are taken in all cases.

This is the main result of this article.
Summarizing, we obtain the analytic expression for the $\ell_1$-norm
recovery threshold by using the two equations in
(\ref{eq:L1deformthreshold}) in a similar form as in the original case.
The difference is that we need to evaluate a double integral in the present 
problem, whereas that in the original case is only single integral
(in the definition of a complementary error function).
When we generalize the deformed problem to the $w$-blockwise model,
we have an expression of the recovery threshold equations with a 
$w$-tuple integral. The problem we previously dealt with \cite{TK2010}
corresponds to the case in which $w=N$, which requires a Monte Carlo
approach to the evaluation of multiple integrals.

\begin{figure} 
\begin{picture}(110,130) 
\put(20,-4){\includegraphics[width=0.39\textwidth]{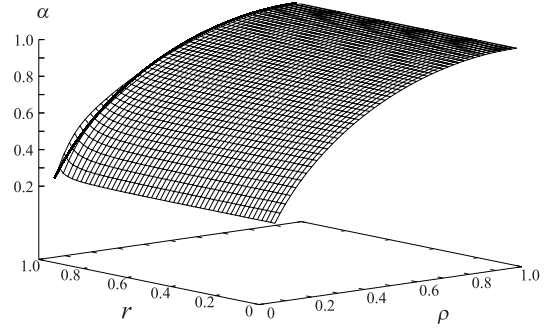}} 
\end{picture} 
\caption{$\ell_1$-norm recovery threshold as function of $\rho$ and $r$.
Areas above and below surface are success and failure regions,
respectively. Thick curve at $r=1$ corresponds to threshold at $r=0$
(i.e. universal threshold),
which is drawn for comparison and for illustrating deviation.
In region of large $r$, a slight deviation from universality is evident.}
\label{fig:figure1} 
\end{figure}

\begin{figure} 
\begin{picture}(110,126)
\put(35,-6){\includegraphics[width=0.34\textwidth]{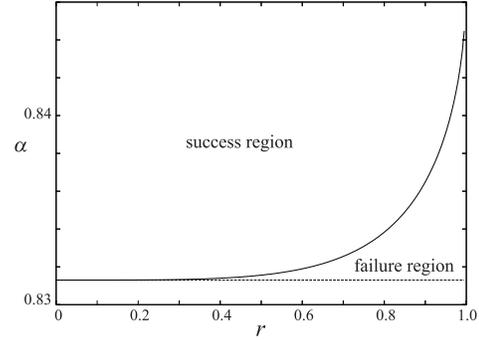}} 
\end{picture}
\caption{Dependence of recovery threshold on correlation parameter
$r$ for $\rho=0.5$. Horizontal line corresponds to universal
threshold value at $\rho=0.5$. For large $r$, 
deviation from universality is apparent.} 
\label{fig:figure2} 
\end{figure}

Using this analytical expression of the recovery threshold for
the deformed problem, we consider the deviation from the universal
threshold. The dependence of recovery threshold $\alpha$ on
two parameters ($\rho$ and $r$) is depicted in Fig. \ref{fig:figure1}.
In the region where $r \approx 1$, a slight deviation from the universal
threshold is evident. The deviation is shown in detail
in Fig. \ref{fig:figure2}, where $\rho=0.5$. In the region of a 
larger $r$, a clear deviation from the universal threshold
(shown by the horizontal line) is evident.

\section{Validation by numerical experiment}
\label{sec:sec6}

Given that the replica method lacks rigorousness, we verified
its results by conducting a numerical experiment of $\ell_1$ recovery.
We used the convex optimization package for MATLAB \cite{GB1,GB2}
and evaluated the recovery threshold. We first prepared a 
square random sensing matrix $\bm F$ with size $N$ and deleted the
rows in $\bm F$ one-by-one until recovery failure occurred,
as determined using $|\bm x_*-\bm x^0| > 10^{-4}$ for recovery
result $\bm x_*$. We recorded the number of remaining rows
(plus one) $P_c=P+1$ at the point of failure. We repeated this
$10^5$ times and then computed the arithmetic average of $P_c/N$,
which we regarded as the value of $\alpha$ at the recovery threshold.

\begin{figure} 
\begin{picture}(110,135)
\put(35,-5){\includegraphics[width=0.34\textwidth]{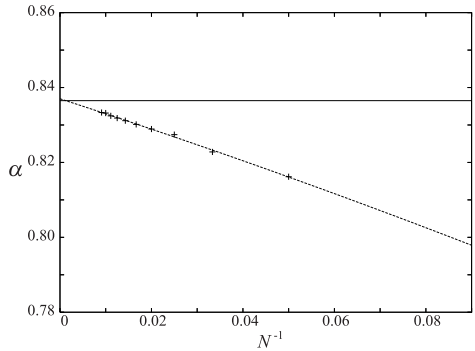}} 
\end{picture} 
\caption{Result of $\ell_1$-norm recovery experiment for $\rho=0.5$
 and $r=0.9$. Number of input signals $N$ was varied from $20$ to $110$
 in steps of $10$. For each $N$, the average was taken over $10^5$ samples.
 Broken curve indicates scaling by quadratic function regression.
 Extrapolated value at $N\rightarrow \infty$ was 
 $0.8370(3)$, while result of statistical mechanics evaluated
 by $N \rightarrow \infty$ limit (horizontal line) was $0.83649...$.
 These results are in excellent agreement.}
\label{fig:figure3} 
\end{figure}

The results are plotted in Fig. \ref{fig:figure3}, in which the
dependence of the threshold value on the dimension of
input signal $N$ is depicted. We also performed scaling analysis 
using quadratic function regression and estimated the value of
the threshold for $N \rightarrow \infty$ limit.
The value of the threshold from extrapolation was 
$0.8370(3)$ for $N \rightarrow \infty$. This value is in excellent
agreement with the result obtained using statistical mechanics
($0.83649...$), which validates our analysis for the 
deformed problem.

\section{Summary and Discussion}
\label{sec:sec7}

We presented a deformed model for $\ell_1$-norm reconstruction, that is, a 
blockwise correlation model that represents the pairwise correlation
in signals. From this model we obtained an analytical expression
for the $\ell_1$-norm recovery threshold by replica heuristics.
Using this expression with a double integral, we evaluated the threshold 
and found a clear deviation from the universal threshold in the region
of strong correlation. A numerical experiment validated the results
obtained with this model. This model enables minute deviations from
the universality of the $\ell_1$ recovery threshold to be traced 
qualitatively by using an analytical expression of the threshold.

We showed that this blockwise model yields a non-universal threshold,
as expected from the rotational symmetry breaking argument.
This suggests that orthogonality of the representation bases
in sensing matrix construction (see \cite{CW2008}) is crucial for 
universality. The importance of orthogonality to universality was
investigated by other researchers in terms of the restricted
isometry property \cite{CW2008,BDDW}, and our results 
support their findings.

The relationship of our analysis to random polytope projection
is of great interest. As we noted in Section \ref{sec:sec2},
for the original model, both geometrical and 
statistical mechanical analyses give the same threshold.
For the deformed problem, like the one we handled using
a blockwise model, the rotational symmetry is broken. It thus appears 
that a ``biased'' projection should be taken into account
in geometrical analysis, where projection group symmetry,
a quotient of rotational group, is generally assumed.
The results presented here should be useful
for obtaining a deeper understanding of the relationship between geometrical
and statistical mechanical analyses.

\section*{Acknowledgments}
This research was supported by KAKENHI (Nos. 22300003 and 22300098)
and the Mitsubishi Foundation (YK).


\end{document}